\documentclass[fdp,fleqn]{w-art}
\usepackage{times}
\usepackage{w-thm}
\usepackage[]{graphicx}

\usepackage{latexsym}
\usepackage{amssymb,amsfonts,amsmath}
\usepackage{indentfirst}
 \usepackage{bbm}

\newcommand{\bsm}{\begin{small}}
\newcommand{\esm}{\end{small}}

\newcommand{\bsubeq}{\begin{subequations}}
\newcommand{\esubeq}{\end{subequations}}

\newcommand{\bfs}{\begin{footnotesize}}
\newcommand{\efs}{\end{footnotesize}}

\newcommand{\bite}{\begin{itemize}}
\newcommand{\eite}{\end{itemize}}

\newcommand {\cC}{{\cal C}}
\newcommand {\cD}{{\cal D}}
\newcommand {\cE}{{\cal E}}

\newcommand {\cJ}{{\cal J}}
\newcommand {\cK}{{\cal K}}

\newcommand {\cM}{{\cal M}}
\newcommand {\cN}{{\cal N}}

\def\a{\alpha}

\def\b{\beta}

\def\d{\delta}

\def\f{\phi}
\def\g{\gamma}
\def\G{\Gamma}

\def\l{\lambda}
\def\m{\mu}

\def\o{\omega}

\def\q{\theta}
\def\r{\rho}
\def\s{\sigma}
\def\t{\tau}

\def\x{\xi}

\def\F{\Phi}

\def\L{\Lambda}
\def\O{\Omega}

\def\S{\Sigma}

\def\rd{{\rm d}}
\def\ri{{\rm i}}

\newcommand{\ve}{\varepsilon}                            
\newcommand{\cDB}{{\bar\cD}}                            

\newcommand{\pa}{\partial}                           
\newcommand{\hf}{\frac12}

\newcommand{\vf}{\varphi}

\newcommand{\be}{\begin{equation}}
\newcommand{\ee}{\end{equation}}
\newcommand{\bea}{\begin{eqnarray}}
\newcommand{\eea}{\end{eqnarray}}
\newcommand{\non}{\nonumber}
\newcommand{\bm}[1]{\mbox{\boldmath$#1$}}

\def\double #1{#1{\hbox{\kern-2pt $#1$}}}

\newcommand{\teb}{{\bar{\theta}}}

\newif\ifdtup

\def\de{{\nabla}}                                         

\def\deb{{\bar{\de}}}

\def\lb{{\bar{\l}}}

\newcommand{\mbS}{{\mathbb S}}

\newcommand{\bfD}{{\bf D}}
\newcommand{\bfDB}{{\bar{\bfD}}}

\begin{document}
\DOIsuffix{theDOIsuffix}
\Volume{55}
\Issue{1}
\Month{01}
\Year{2007}
\pagespan{3}{}
\Receiveddate{2012}
\Reviseddate{2012}
\Accepteddate{2012}
\Dateposted{2012}
\keywords{Extended Supersymmetry, AdS, Supergravity Models, Superspaces.}

\title[Topics in 3D $\bm{\cN = 2}$ AdS supergravity in superspace]{Topics in 3D $\bm{\cN = 2}$
 AdS supergravity in superspace}

\author[Gabriele Tartaglino-Mazzucchelli]{Gabriele Tartaglino-Mazzucchelli\inst{1,}%
  \footnote{E-mail: \textsf{gabriele.tartaglino-mazzucchelli@physics.uu.se}}}
\address[\inst{1}]{Theoretical Physics, Department of Physics and Astronomy,
Uppsala University \\ 
Box 516, SE-751 20 Uppsala, Sweden}
\begin{abstract}
We review some recent results on the construction in superspace
 of 3D $\cN=2$ AdS supergravities 
 and on the formulation of rigid supersymmetric theories in (1,1) and (2,0) AdS superspaces.
\end{abstract}
\maketitle

\section{Introduction}
\setcounter{equation}{0}

Recently there has been a renewed interest in 3D supersymmetric theories.
On the one hand this was generated by the 
new insights achieved in the study of M2-brane dynamics and 3D superconformal Chern-Simons
theories \cite{BL,G,ABJM}. 
Such results triggered  a large field of investigations on AdS$_4$/CFT$_3$ dualities.

On the other hand,
new insights into the dynamics of 3D massive gravity theories have been achieved. 
The use of AdS$_3$/CFT$_2$ duality has allowed 
a microscopic derivation of the BTZ black hole entropy using
Topological-Massive-Gravity (TMG) \cite{Li-Song-Strominger-08}.
After that, new classes of higher-derivative, but unitary, 3D gravities,
called New-Massive-Gravity (NMG) and
Generalized-Massive-Gravity (GMG), were constructed
\cite{Bergshoeff-Hohm-Townsend-09}.
These theories were extended at the non-linear level to $\cN=1$ supergravity 
\cite{Andringa-Bergshoeff-deRoo-Hohm-Sezgin-Townsend-09,Bergshoeff-Hohm-Rosseel-Sezgin-Townsend-10}
and,  in the linearized approximation, to $\cN\geq 2$ 
\cite{Bergshoeff-Hohm-Rosseel-Townsend-10}.
A fully non-linear description of the supergravity extension is difficult due to the 
higher derivative terms.
A crucial role is played by the three-dimensional AdS space which  represents 
a maximally symmetric solution in these models, making them of interest for AdS$_3$/CFT$_2$.

The use of superspace techniques for $\cN$-extended supergravity may give insights into the 
generalization of the previous results.
Surprisingly this was not fully developed in the past.
The $\cN=1$ case was studied in \cite{Howe-Tucker-1977,BG}.
The  $\cN\geq 2$ was sketched in \cite{HIPT}
(more results for the $\cN=8$ case were given in \cite{Howe-Sezgin-04}).
To fill this gap, in collaboration with S.~M.~Kuzenko and U.~Lindstr\"om,
in \cite{KLT-M} we developed the superspace description of
$\cN$-extended conformal supergravity\footnote{Independently, 
in \cite{CGN} and \cite{GH} the superspace geometry of respectively $\cN=8$ and $\cN=16$ 
supergravity was studied.}
and, extending the superconformal results of \cite{KPT-MvU},
we provided formalisms to study
general supergravity-matter systems with $\cN\le 4$.

The development of a superspace approach to study field theories in curved spaces
has received a renewed attention (see \cite{BK_AdS_supercurrent,BX,BKsigma,AJKL,FS}).
For example, rigid supersymmetric sigma-models in 4D AdS have revealed new restrictions
on the target space geometry and on the structure of the allowed supercurrent multiplets.
New superspace techniques for rigid supersymmetric theories
on curved background have clear applications if one is interested to lift
 off-shell theories from flat to curved backgrounds \cite{FS}.
Such problems have recently arisen
in  studying the partition function of gauge
theories on nontrivial 3D-4D, constant-curvature backgrounds 
(mostly spheres) when computing
observables such as   expectation values of Wilson loops  
and superconformal indices by using localization techniques 
\cite{Localization}.

The 3D constant curvature spaces present interesting features.
For example, long ago,  Ach\'ucarro and Townsend discovered that 3D
$\cN$-extended anti-de Sitter (AdS) supergravity exists in several incarnations,
called (p,q) supergravities \cite{AT}.
The two non-negative integers $p \geq q$ are such that $\cN=p+q$
and they classify the 
in-equivalent isometry supergroups  $\rm OSp(p|2;{\mathbb R}) \times OSp(q|2;{\mathbb R})$
of the AdS backgrounds.

In \cite{KT-M-09-2011}, using the results of \cite{KLT-M},  we deepened the study of 
3D $\cN=2$ AdS supergravities in superspace. 
We presented three superfield formulations 
for $\cN=2$ supergravity that allow for well defined  cosmological terms and supersymmetric 
AdS solutions. 
We classified the consistent supercurrent multiplets
in flat, (1,1) AdS and (2,0) AdS superspaces.
We proved that both (1,1) and (2,0) AdS are conformally flat superspaces.
Furthermore, we elaborated on rigid supersymmetric theories in (1,1) and 
(2,0) AdS superspaces.

In this report we review some of the results obtained in \cite{KLT-M,KT-M-09-2011}
for the 3D $\cN=2$ case.
In section 2, we review the superspace formulation of 3D $\cN=2$ conformal supergravity.
Section 3 is devoted to the classification of supergravity compensators and cosmological terms
that give rise to (1,1) and (2,0) AdS supergravities.
In section 4 we present some results concerning the formulation of rigid supersymmetric
models in (1,1) and (2,0) AdS superspaces.

\section{3D {\bm{\cN = 2}} conformal supergravity in superspace}
\setcounter{equation}{0}
\label{SCG}

We start by describing the superspace formulation of 3D $\cN=2$ off-shell conformal supergravity
originally presented as part of the general $\cN$ analysis of \cite{KLT-M} 
and further developed in \cite{KT-M-09-2011}.

Consider a curved 3D $\cN=2$ superspace  $\cM^{3|8}$ parametrized by
local bosonic ($x$) and fermionic ($\q, \bar \q$)
coordinates  $z^{{M}}=(x^{m},\q^{\mu},{\bar \q}_{{\mu}})$,
where $m=0,1,2$, $\mu=1,2$.
The Grassmann variables $\q^{\mu} $ and $\teb_{{\mu}}$
are related to each other by complex conjugation
$\overline{\q^{\mu}}=\teb^{{\mu}}$.
The tangent-space group is chosen to be ${\rm SL}(2,{\mathbb{R}})\times {\rm U(1)}_R$
and the superspace covariant derivatives
$\cD_{{A}} =(\cD_{{a}}, \cD_{{\a}},\cDB^\a)$
have the form
\bea
\cD_{{A}}=E_{{A}}
+\O_{{A}}
+\ri \,\F_{{A}}\cJ~.
\label{CovDev}
\eea
Here $E_{{A}}=E_{{A}}{}^{{M}}(z) \pa/\pa z^{{M}}$
is the supervielbein with $\pa_M=\pa/\pa z^M$;
$\cM_{bc}$ and $\O_{A}{}^{bc}$ are the Lorentz generators and connection respectively
(antisymmetric in $b,\,c$);
$\cJ$ and $\F_{A}$ are respectively the ${\rm U(1)}_R$ generator and connection.
The generators of SL(2,$\mathbb{R})\times {\rm U(1)}_R$
act on the covariant derivatives as follows:\footnote{Note that  the (anti)symmetrization of $n$ 
indices is defined to include a factor of $(n!)^{-1}$.}
\bea
{[}\cJ,\cD_{\a}{]}
=\cD_{\a}~,
\quad
{[}\cJ,\cD_a{]}=0~,
\qquad
{[}\cM_{\a\b},\cD_{\g}{]}
=\ve_{\g(\a}\cD_{\b)}~,\quad
{[}\cM_{ab},\cD_c{]}=2\eta_{c[a}\cD_{b]}~.
\label{generators}
\eea
Here, 
$\cM_{\a\b}=(\g^a)_{\a\b}\cM_{a}$ are the Lorentz generators 
with $(\g_a)_{\a\b}$ the symmetric and real gamma-matrices,
$\ve_{\a\b}$ is the antisymmetric SL(2,$\mathbb{R}$) invariant
and $\cM_{a}=\hf\ve_{abc}\cM^{bc}$ being
$\ve_{abc}$ ($\ve_{012}=-1$) the Levi-Civita tensor
(see \cite{KLT-M,KT-M-09-2011} for more details on our 3D notations and conventions).

The supergravity gauge group is generated by local transformations
of the form
\bea
&\d_K \cD_{{A}} = [K  , \cD_{{A}}]~,
\quad
\d_K U = K\, U~,
\qquad K = K^{{C}}(z) \cD_{{C}} +\hf K^{ c d }(z) \cM_{c d}
+\ri \, \t (z) \cJ  ~,
\label{tau}
\eea
with the gauge parameters
obeying natural reality conditions, but otherwise  arbitrary.
In (\ref{tau}) we have included the transformation rule for a tensor superfield
$U(z)$, with its indices suppressed.

If one imposes conventional constraints  \cite{HIPT}
and solves the the Bianchi identities \cite{KLT-M},
the covariant derivatives algebra turn out to obey the 
(anti)commutation relations ($\cD_{\a\b}=(\g^a)_{\a\b}\cD_a$)
\bsubeq
\bea
\{\cD_\a,\cD_\b\}
&=&
-4\bar{R}\cM_{\a\b}
~,\qquad\qquad
\{\cDB_\a,\cDB_\b\}
~=~
4{R}\cM_{\a\b}~,
\label{N=2-alg-1}
\\
\{\cD_\a,\cDB_\b\}
&=&
-2\ri\cD_{\a\b}
-2\cC_{\a\b}\cJ
-\ri\ve_{\a\b} {\mathbb S} \cJ
+\ri {\mathbb S} \cM_{\a\b}
-2\ve_{\a\b}\cC^{\g\d}\cM_{\g\d} ~,
\label{dim-1-algebra}
\\
{[}\cD_{\a\b},\cD_\g{]}
&=&
-\ri\ve_{\g(\a}\cC_{\b)\d}\cD^{\d}
+\ri\cC_{\g(\a}\cD_{\b)}
-\hf\ve_{\g(\a} {\mathbb  S}\cD_{\b)}
-2\ri\ve_{\g(\a}\bar{R}\cDB_{\b)}
\non\\
&&
+~{\rm curvature~terms}~.
\label{dim-3/2-algebra}
\eea
\esubeq
The algebra is parametrized by three dimension-1 torsion superfields:
a real scalar $\mathbb S$, a
complex scalar $R$ and its conjugate $\bar{R}$,  and a real vector $\cC_a$
($\cC_{\a\b}:=(\g^a)_{\a\b}\cC_a$).
The superfields $\mathbb S$ and $\cC_a$  are neutral under the ${\rm U(1)}_R$ group, 
while the ${\rm U(1)}_R$ charge of $R$ is $-2$, 
 $\cJ R=-2R$ and $\cJ \bar{R}=2\bar{R}$.
The torsion superfields obey differential constraints implied by the Bianchi identities.
At dimension-3/2 these are
\bea
&\cDB_\a R=0~,
\qquad
\cD_{\a}\cC_{\b\g}
=
\ri C_{\a\b\g}
+\frac{\ri }{ 3}\ve_{\a(\b}\Big(
\ri \cDB_{\g)}\bar{R}
-\cD_{\g)}{\mathbb S}
\Big)
~,
\label{N2SG-dim-3/2-constr}
\eea
together with their complex conjugates.
These imply at dimension-2 the following descendant equation
\bea
(\cD^2
-4\bar{R})\mathbb S
=(\cDB^2
-4R)\mathbb S
=0
~,\qquad\qquad
\cD^2:=\cD^\g\cD_\g
~,\qquad
\cDB^2=\cDB_\g\cD^\g
~.
\label{4.16--}
\eea
The constraints tell us that $R$ and  $\mathbb S$ are respectively
chiral and real linear superfields.
It is not surprizing  that the 3D $\cN=2$ geometry has the resemblance of a dimensionally 
reduced version of 4D $\cN=1$ conformal supergravity in superspace
(for reviews on 4D $\cN=1$ supergravity see \cite{GGRS,Ideas}).

The fact that the supergeometry introduced corresponds to 3D $\cN=2$ conformal supergravity,
relies on the fact that the algebra (\ref{N=2-alg-1})--(\ref{dim-3/2-algebra})
and the Bianchi identities 
are invariant under super-Weyl transformations
of the covariant derivatives\footnote{We omit the transformations 
$\d\cD_a$ which are induced by the one of the spinor covariant derivatives
$\d_\s\cD_\a$  \cite{KLT-M}.}
\bsubeq\label{sW-general}
\bea
&\d_\s\cD_\a=\hf\s\cD_\a+(\cD^{\g}\s)\cM_{\g\a}-(\cD_{\a }\s)\cJ~,
\eea
\esubeq
where the scalar superfield $\s$ is real and unconstrained.
The dimension-1 torsion  components 
transform as
\bea
\d_\s R=
\s R
+\frac{1}{4}(\cDB^2\s)
~,
\quad
\d_\s{\mathbb S}=\s{\mathbb S}
+\ri(\cD^\g\cDB_{\g}\s)
~,
\quad
\d_\s\cC_{a}=\s
\cC_{a}
+\frac{1}{8}(\g_a)^{\g\d}([\cD_{\g},\cDB_{\d}]\s)
~.~~~~~~
\label{sW-C-1}
\eea
It can be proved that, by using super-Weyl transformations, many components of the
supergravity multiplet embedded in the geometry are gauged away. The remaining 
component fields are
the vielbein $e_a{}^m$, the gravitini $\Psi_a{}^\mu$ and the U(1)$_R$ connection
$A_a$ with no auxiliary fields and no Weyl tensor. 
These are exactly the field components of the 3D $\cN=2$ Weyl multiplet of conformal 
supergravity (see for example \cite{RvN}).

\section{3D ${\bm \cN=2}$ AdS supergravities}

We have reviewed the geometric description of 3D $\cN=2$ conformal supergravity in 
superspace.
To generate the different 3D $\cN=2$  AdS supergravities we first need to study 
the classes of scalar multiplets that can be used as conformal compensators.
Depending on the compensator chosen, the different cosmological terms give rise to 
the two inequivalent 3D $\cN=2$ AdS superspaces as solution of the supergravity equation of 
motions.
Below we illustrate these steps.

As pointed out above, 3D $\cN=2$ conformal supergravity is analogue to the 4D $\cN=1$
one.
In fact, there are three different natural types 
of scalar multiplets that can be used as conformal compensators 
to generate Poincar\'e supergravity. As in 4D \cite{GGRS,Ideas}, these are:\\
(i) a complex  chiral superfield $\F$ satisfying
\bea
&\cDB_\a\F=0~,\qquad
\d_\s\F=\hf\s\F~,
\qquad\cJ\F=-\hf\F
~;
\eea
(ii) a real  linear superfield $G$ such that
\bea
(\cDB^2-4R){\mathbb G}=0~, \qquad
\overline{({\mathbb G})}={\mathbb G}
~,
\qquad
\d_\s{\mathbb G}=\s{\mathbb G}
\qquad
\cJ{\mathbb G}=0
~;
\label{real-linear}
\eea 
(iii) a complex linear superfield $\S$ that obeys the conditions
\bea
(\cDB^2-4R)\S=0~,\qquad
\d_\s\S=w\s\S~,
\qquad
 \cJ\S=(1-w)\S
 ~.
\eea
The constant parameter $w$ is real and constrained to be different from zero and one,
 $w\ne0,1$, to ensure
that $\S$ has nontrivial super-Weyl and U(1)$_R$ transformations.
The three choices of compensators respectively provide the 3D analogous of 4D $\cN=1$:
(i) old-minimal, 
(ii) new-minimal
and (iii) non-minimal supergravities (look at \cite{Ideas} for a detailed list of references
on the 4D $\cN=1$ supergravities).

To describe 3D AdS supergravity one has to introduce cosmological terms that depend on the
choice of the compensators.
In the case (i), the supergravity action is\footnote{Here $E$ is the full superspace density
with $E^{-1}={\rm Ber}(E_A{}^M)$ 
and $\cE$ is the chiral density.}
\bea
S^{(1,1)}_{\rm{AdS}} =-4\int {\rm d}^3x {\rm d}^4 \q 
\,E\,\bar \F \F
+{\m}\int {\rm d}^3x {\rm d}^2 \q\,\cE\, \F^4
+{\bar \m}\int {\rm d}^3x  {\rm d}^2 {\bar \q}\,\bar \cE\,\bar \F^4~,
\label{3.11}
\eea
where $\mu$ is a complex constant parameter that plays the role of the cosmological constant.
One can prove that the following equations hold on-shell
\bea
&\cC_a=\mbS=0~,
\qquad
R=\mu
~.
\label{1-1-eqom}
\eea
Denoting with $\de_A=(\de_a,\de_\a,\deb_\a)$ the on-shell covariant derivatives, 
their algebra turn out to be
\bsubeq \label{11AdSsuperspace}
\bea
&\{\de_\a,\de_\b\}
=
-4\bar{\mu}\cM_{\a\b}
~,\qquad
\{\de_\a,\deb_\b\}
=
-2\ri\de_{\a\b}~, 
\label{AdS_(1,1)_algebra_1}
\\
&{[}\de_{\a\b},\de_\g{]}
=
-2\ri\bar{\mu}\ve_{\g(\a}\deb_{\b)}
~,\qquad
{[}\de_a,\de_b]{}
= -4 \bar{\mu}\mu \cM_{ab} ~.
\label{AdS_(1,1)_algebra_2}
\eea
\esubeq
According to the classification of \cite{AT},
the latter describes the 3D (1,1) AdS superspace
and the (\ref{3.11}) theory describes AdS supergravity.
Note that on-shell the U(1)$_R$ generator disappear from the algebra.

The previous theory is the analogue of 4D $\cN=1$ old-minimal supergravity with a cosmological 
constant term \cite{Ideas}.
Recently it was proved that an equivalent description can be achieved
in non-minimal supergravity \cite{BKdual}.
The same is true in 3D \cite{KT-M-09-2011}.
The basic idea is to notice that, given a general scalar complex superfield such that
$\d_\s \G= w\s\G$, $\cJ\G=(1-w)\G$, then it holds:
$\d_\s \big((\cDB^2-4R)\G\big)=(1+w) \s(\cDB^2-4R)\G$.
The complex linear  $\S$ is an example of such a superfield
and the latter equation tells us that the linear constraint is super-Weyl invariant
as it should be to make $\S$ a proper conformal compensator.
On the other hand, if $w=-1$,
then $(\cDB^2-4R)\G$ is trivially super-Weyl invariant and 
the linear constraint can be consistently deformed.
To describe AdS supergravity, one can use a conformal compensator
such that  $\d_\s \G= -\s\G$, $\cJ\G=2\G$, and satisfying the \emph{improved} linear 
constraint\footnote{In global 4D $\cN=1$ supersymmetry, constraints  
of this form were introduced by Deo and Gates \cite{DG85}.
In the context of 4D supergravity, such constraints have recently been used in \cite{KTyler}
and then in \cite{BKdual}.}
\bea
-\frac{1}{4} (\bar \cD^2 - 4 R) \Gamma = \m ={\rm const}~.
\eea
A dual formulation of the 3D, (1,1) AdS theory 
(\ref{3.11}) is described by the first-order action \cite{KT-M-09-2011}
\bea
S^{(1,1)}_{\rm{AdS}} = -2 \int {\rm d}^3x {\rm d}^4 \q 
\,E\,
{ 
{ (\bar \G \, \G)} 
}^{-1/2}~.
\label{non-minimal-AdS}
\eea
In fact, the equations of motion arising from (\ref{non-minimal-AdS}) are exactly
(\ref{1-1-eqom}). The on-shell algebra
is then the one of (1,1) AdS superspace (\ref{AdS_(1,1)_algebra_1})--(\ref{AdS_(1,1)_algebra_2}).

We are left with the case (ii) of a real linear compensator.
In 4D it is known that no cosmological constant term is allowed in new minimal supergravity.
However, in 3D the situation is different and more interesting.
Note that the constraints (\ref{real-linear}) describe 
a 3D $\cN=2$ Abelian vector multiplet.
Instead of the field strength ${\mathbb G}$, we can be use the real unconstrained
prepotential superfield $G$ such that: ${\mathbb G}=\ri\cD^\a\cDB_\a G$, $\d_\s G=\cJ G=0$.
This is defined up to gauge transformations $\d G=(\l+\lb)$ generated by a chiral superfield $\l$
such that: $\cDB_\a\l=0$ and $\d_\s\l=\cJ\l=0$.
The super-Weyl invariant, AdS supergravity action for a vector multiplet compensator
has the following form \cite{KT-M-09-2011} 
($\r$ is a real coupling constant)
\bea
S^{(2,0)}_{\text{AdS}}=\int {\rm d}^3x {\rm d}^4 \q 
\,E\,\big(L_{\text{IT}}+L_{\text{CS}}\big) ~,\qquad
L_{\text{IT}}  = 4\big( \mathbb G \ln \mathbb G - G{\mathbb S}\big)
~,
\quad
L_{\text{CS}}=2\r\, G \mathbb G
~.
\label{Type-II-AdS}
\eea
Here $L_{\text{IT}}$ is a supergravity extension the improved tensor multiplet Lagrangian
\cite{HitchinKLR,deWR}
and is the 3D analogue of the 4D $\cN=1$ new-minimal supergravity Lagrangian
\cite{GGRS,Ideas}.
The Chern-Simons Lagrangian $L_{\text{CS}}$ is the 3D novelty and it represents a cosmological
term. 
The equations of motion of (\ref{Type-II-AdS}) are
\bea
&\cC_a=R=0~,\qquad
\mathbb S=\r
~.
\eea
Denoting with $\bfD_A=(\bfD_a,\bfD_\a,\bfDB_\a)$ the on-shell covariant derivatives, their algebra 
is given by
\bsubeq \label{20AdSsuperspace}
\bea
&\{\bfD_\a,\bfD_\b\}
=0
~,\qquad
\{\bfD_\a,\bfDB_\b\}
=
-2\ri\bfD_{\a\b}
-\ri \r \ve_{\a\b} \cJ
+\ri\r \cM_{\a\b} ~, 
\label{AdS_(2,0)_algebra_1}
\\
&{[}\bfD_{\a\b},\bfD_\g{]}
=
-\hf \r \ve_{\g(\a} \bfD_{\b)}~,\qquad
{[}\bfD_a,\bfD_b{]} = - \frac{1}{4} \r^2 \cM_{ab}~.
\label{AdS_(2,0)_algebra_2}
\eea
\esubeq
According to the classification of \cite{AT},
the latter describes the 3D (2,0) AdS superspace
and (\ref{Type-II-AdS})  describes an AdS supergravity theory.
This is inequivalent to the (1,1) supergravity described by
the chiral and non-minimal compensators. Note that in the (2,0) case the 
U(1)$_R$ generator remains part of the algebra.
The (1,1) and (2,0) AdS superspaces
are characterized by quite different geometrical features that affect the matter systems
that can be consistently formulated in these geometries. In what follows, we turn our attention
to describing some matter systems in the two 3D AdS superspaces and to pointing out some
of their differences.


\section{Matter couplings in AdS superspaces}
\setcounter{equation}{0}

We turn to considering 
rigid supersymmetric field theories in (1,1) AdS superspace.
These are theories invariant under the isometry  transformations of the (1,1)  AdS
geometry.
The isometries are generated by Killing vector fields, 
$\L=\l^a\de_a+\l^\a\de_\a+\lb_\a\deb^\a$, 
which, combined with an appropriate Lorentz transformation,
leave invariant the covariant derivatives:
$\big{[}\L+\hf\o^{ab}\cM_{ab},\de_C\big{]}=0$ \cite{KT-M-09-2011}.
It can be shown that the (1,1) AdS Killing vector fields generate the supergroup 
$\rm OSp(1|2;{\mathbb R}) \times OSp(1|2;{\mathbb R})$.

Matter couplings in  (1,1) AdS superspace are very similar to those in 4D $\cN=1$ AdS 
\cite{AJKL,FS,BKsigma}, and they are more restrictive than their flat counterparts.  
As a nontrivial example, here we consider the most general 
supersymmetric nonlinear $\s$-model in (1,1) AdS superspace described by the action
\bea
S = \int {\rm d}^3x {\rm d}^4 \q 
\,E\,
\cK(\vf^I,\bar{\vf}^{\bar{J}})~.
\label{9.15}
\eea
Here $\vf^I$ are chiral superfields, ${\bar \nabla}_\a \vf^I =0$,
and at the same time local complex coordinates of a complex  manifold $\cM$.
The action is invariant under (1,1) AdS isometry transformations
$\d \vf^I = \L\vf^I $.

Unlike in the Minkowski case, the action does not possess K\"ahler
invariance. This relies in the relation
\bea\label{2.6}
\int \rd^3x\, \rd^4\theta \, E\, F (\vf)= \int \rd^3x\, \rd^2\theta \, \cE\, \mu F (\vf)  \neq 0~,
\eea
which relates every chiral integral of a holomorphic function to a full superspace integral.
It turns out that, because of (\ref{2.6}), the Lagrangian $\cK$ in  (\ref{9.15}) should be  a globally 
defined function on the K\"ahler target space $\cM$. 
This implies that the K\"ahler two-form, 
 $ \O=2\ri \,g_{I \bar J} \, \rd \vf^I \wedge \rd \bar \vf^{\bar J}$,  associated with 
the K\"ahler metric $g_{I \bar J} := \partial_I \partial_{\bar J} \cK$, 
is exact and hence  $\cM$ is necessarily non-compact exactly as the 4D AdS case
\cite{AJKL,FS,BKsigma}. 
The $\s$-model couplings in (1,1) AdS are more restrictive than in the 3D Minkowski case.

What is the situation in the (2,0) AdS superspace?
The  Killing vector fields,
$\t=\t^a\bfD_a+\t^\a\bfD_\a+\bar{\t}_\a\bfDB^\a$, generating the
 isometries of (2,0) AdS superspace, 
 obey the equation
$\big{[}\t+\ri t\cJ+\hf t^{bc}\cM_{bc},\bfD_A\big{]}=0$
where $t$ and $t^{bc}$ are constrained U(1)$_R$ and Lorentz parameters \cite{KT-M-09-2011}.
The (2,0) AdS Killing vector fields  prove to generate the supergroup 
$\rm OSp(2|2;{\mathbb R}) \times Sp(2,{\mathbb R}) $.
Note that, due to the U(1) subgroup,
matter couplings in  (2,0) AdS superspace differ from those in the (1,1) case.
In fact, only $R$-invariant actions can be consistently defined in 
 (2,0) AdS superspace. Moreover, since in (2,0) superspace the chiral curvature is zero,
$R=0$, chiral integrals cannot be rewritten as a full superspace integral in contrast with the
(1,1) case spelled out by eq. (\ref{2.6}). This indicates that holomorphicity may still be an important 
ingredient in studying the dynamics in (2,0) AdS similarly to the flat case.
 As an example, consider the action
\begin{align}
S = \int {\rm d}^3x {\rm d}^4 \q \,E\,
K(\phi^I, \bar \phi^{\bar J})
+\Big{\{} \int {\rm d}^3x {\rm d}^2 \q \,\cE\,
W(\phi^I)
\,+\,{\rm c.c.}~\Big{\}} ~.
\label{10.11}
\end{align}
Here $\f^I$ are chiral superfields, $\bar {\bf D}_\a \f^I=0$,
with U(1)$_R$ charges $r_I$:
$\cJ \f^I= -r_I \f^I $
(no sum over I).
For $R$-invariance, 
the K\"ahler potential $K(\f , \bar \f )$
and the superpotential $W (\f )$ should obey:
\bea
\sum_I r_I \f^I K_I =\sum_{\bar{I}} r_I \bar \f{}^{\bar{I}} K_{\bar{I}}~,
\qquad
\sum_I r_I \f^I W_I=2W
\label{10.17a} ~.
\eea
The action is invariant under the isometry transformations
$\d \phi^I=
\big(\t+\ri t\cJ\big)\phi^I$.
An important class of $\s$-models  in (2,0) AdS superspace
is specified by the conditions $r_I=0$ and $W(\vf)=0$. In this case no restrictions on 
the K\"ahler target space occur and, unlike the (1,1) case,
compact target spaces are allowed.

Let us conclude by describing
a system of self-interacting 
Abelian vector multiplets described by 
real linear field strengths ${\mathbb F}^i $, with $i=1,\dots, n$.
A general gauge invariant action in (2,0) AdS is
\bea
S=\int {\rm d}^3x {\rm d}^4 \q 
\,E\,\Big\{  L({\mathbb F}^i) 
+ \hf m_{ij} F^i {\mathbb F}^j 
+ \x_{i}F^i\Big\}~,
\label{general-abelian-2-0}
\eea
with $ m_{ij} =m_{ji} =(m_{ij})^*$ and $\x_i$ being Chern-Simons and Fayet-Iliopoulos coupling
constants 
respectively.
Here $F^i$ is the gauge prepotential for ${\mathbb F}^i$
and $L$ is an arbitrary real function of ${\mathbb F}^i $.
The scalar superfields $F^i$ and ${\mathbb F}^i$ have isometry transformations
$\d F^i=\t F^i$,
$\d {\mathbb F}^i=\t {\mathbb F}^i$.

It is interesting to note that 
the very same action (\ref{general-abelian-2-0}) is well defined in (1,1) AdS only if the 
Fayet-Iliopoulos term is not present, $\x_i\equiv 0$. In fact, these are gauge invariant only
in (2,0) AdS where $R=0$.


\begin{acknowledgement}
We thank the organizers of the XVII European Workshop on String Theory 2011
for the opportunity to report on these results.
We are grateful to S.~M.~Kuzenko and U.~Lindstr\"om for collaborations and discussions.
This work is supported by the European 
Commission, Marie Curie IEF under contract No.
PIEF-GA-2009-236454.  
\end{acknowledgement}

\end{document}